\documentclass[preprint,showpacs,showkeys,preprintnumbers,amssymb,aps]{revtex4}

\usepackage{graphicx}
\usepackage{dcolumn}
\usepackage{bm}

\begin{document}


\title{Skyrme Hartree-Fock Calculations for the Alpha Decay Q Values of
Super-Heavy Nuclei}

\author{S. Typel}
\altaffiliation[Present address: ]{Gesellschaft f\"{u}r 
Schwerionenforschung mbH, Planckstra\ss{}e 1, 64291 Darmstadt, Germany}
\author{B. A. Brown}
\affiliation{
Department of Physics and Astronomy and
National Superconducting Cyclotron Laboratory,
Michigan State University, 
East Lansing, Michigan 48824-1321
}

\date{\today}

\begin{abstract}
{Hartree-Fock calculations with the SKX Skyrme interaction are
carried out to obtain
alpha-decay Q values for deformed nuclei above $^{208}$Pb
assuming axial symmetry. The results for
even-even nuclei are compared with experiment and with previous
calculations. Predictions are made for alpha-decay Q values and
half-lives of even-even super-heavy nuclei. The results are also
compared for the recently discovered odd-even chain starting at
$  Z=112  $ and $  N=165  $.}
\end{abstract}

\pacs{23.60.+e, 21.10.Dr, 21.60.Jz, 27.90.+b}
\keywords{alpha decay, super-heavy elements,
Skyrme Hartree-Fock calculations, deformed nuclei}

\maketitle

\section{Introduction}

The existence and decay properties of super-heavy nuclei are one
of the most fundamental problems in nuclear physics \cite{arm,hof}.
There are now new data which
confirm the existence of $  Z  $ = 111 and 112 and their
connection to lighter decay chains \cite{z112}. The first
data for $  Z=114  $ and
$  Z=116  $ also exist \cite{z116}, 
with suggested $  A=288  $ and $  A=292  $,
respectively, but the $  A  $ values are not certain since the
connection to lighter nuclei is not known.
Theoretical models for super-heavy nuclei
have evolved from the
macroscopic-microscopic models such as the finite-range droplet model
with shell corrections \cite{Mol95},
to fully microscopic deformed Hartree-Fock (HF) models
such as those presented in \cite{Cwi99} and \cite{Gor01}.
In addition to their intrinsic many-body nuclear structure importance,
theoretical models for the prediction of the decay properties of the
super-heavy nuclei are important when designing experiments
since the techniques used will depend on the half-life and decay mode.

In this paper we present a new set of Hartree-Fock results for
alpha-decay Q-values of super-heavy nuclei. A global formula is used to
calculate the half-lives. Our calculations are based upon a new
computational program for solving the axial-symmetric HF equations and
the new Skyrme interaction SKX \cite{Bro98}. The reason for exploring
results with another Skryme interaction beyond those used
previously in \cite{Cwi99}
and \cite{Gor01} is that the alpha decay Q-value systematics are 
sensitive to the spherical and deformed shell-effects 
which depend upon the
underlying parameters of the hamiltonians. There are several modern
Skyrme parameter sets available, each of them determined with a
different weighting and emphasis on the existing nuclear structure
properties.

The SLy4 parameters \cite{sly4} used in \cite{Cwi99} and the MSk7 
\cite{Gor01} parameters take into account
overall spacing of the single-particle states such as those in 
$^{208}$Pb.
(In particular, it is common to constrain the Skyrme parameters to give
an effective mass of unity which is required by the observed level
spacing.) However, the SKX interaction \cite{Bro98} explicitly 
incorporates
most of the observed single-particle levels in $^{208}$Pb into the data
set which was used to determine the parameter values. The 
single-particle
energies for some proton and neutron particle states above the fermi
surface in $^{208}$Pb are compared in Table I. All of the calculations
have some disagreement with experiment, however, SKX has the best
overall agreement with experiment. It is important to explore the
model-dependences of the HF results for the binding-energies of nuclei
above $^{208}$Pb.

\begin{figure}[ht]
\centerline{\includegraphics[scale=0.6]{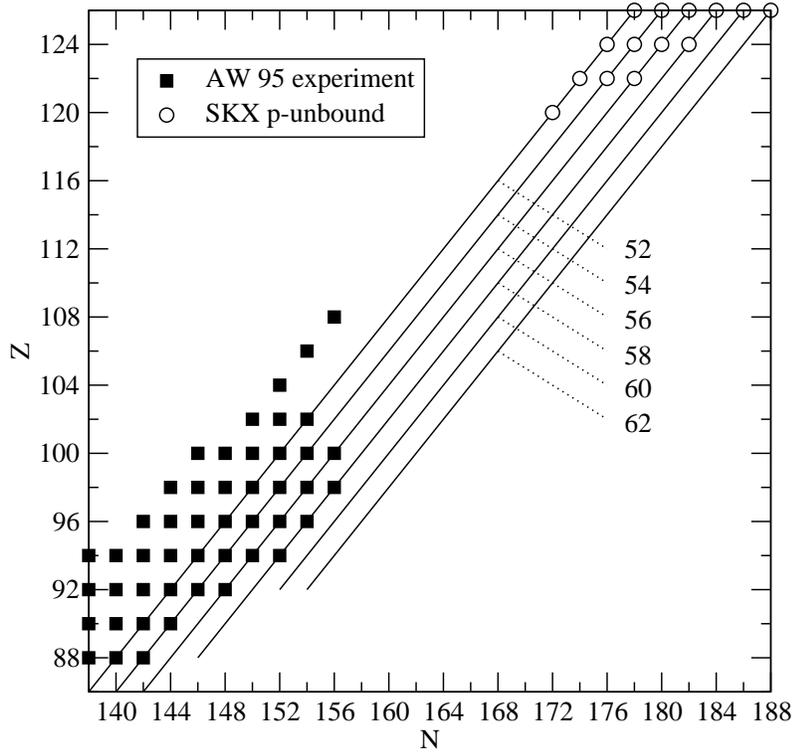}}
\caption{\label{fig1}
Chart of the nuclides for even-even nuclei with
$  N \geq 138  $ and $  Z \geq 86  $.
Solid squares denote nuclei with experimentally known
masses. Nuclei discussed in this paper are located on
the solid lines with constant $N-Z$ (indicated by the numbers). 
Open circles indicate
proton-unbound nuclei in the SKX parametrization.}
\end{figure}

In Sec.\ \ref{sec2} we will briefly discuss the computation method
for the binding energies and single-particle spectra
for axial symmetric nuclei in the  HF approximation.
In Sec.\ \ref{sec3} the results for alpha
decay Q-values and lifetimes will be presented and discussed.
The region of the chart of nuclides covered by our calculations is
shown in Fig.\ \ref{fig1}. 
This includes the region just above $^{208}$Pb where 
the alpha-decay Q-values are measured, and extends up to the assumed
spherical magic numbers of $  Z=126  $ and $  N=184  $ for the 
super-heavy
predictions.

\section{Method of calculation}
\label{sec2}

There are several methods available for solving the
Schr\"{o}dinger equation for the single particle wave functions
$  \phi_{i}  $
in a Skyrme-type potential with axial symmetry.
Often a cylindrical coordinate system is chosen and the wave
functions are expanded in a deformed harmonic oscillator basis with
carefully adjusted oscillator strengths. Then the Hamiltonian
is diagonalized in an appropriately truncated space of basis states.
Here, we solve the Schr\"{o}dinger
equation for $  \phi_{i}  $ in coordinate space with a spherical basis 
for the
angular part of the wave function. In this approach there is a smooth
transition to the case of spherical nuclei where the wave function
simplifies considerably.

The single-particle wave functions
of protons ($  q=+1  $) and neutrons ($  q=-1  $)
in a deformed nucleus with axial symmetry are specified by three 
quantum numbers:
the parity $  \pi=\pm 1  $, the principal quantum 
number $  n=1,2,\dots  $,
and the projection of the total angular momentum  on the symmetry-axis
$  \Omega=\pm \frac{1}{2},\pm \frac{3}{2}, \dots  $. We expand the
wave functions in coordinate space
\begin{equation} \label{phi}
 \phi_{q\pi n \Omega}(\vec{r}) =
 \frac{1}{r} \sum_{\kappa} f_{q \pi n \kappa \Omega}(r)
 {\cal Y}_{\kappa \Omega}(\hat{r})       
\end{equation}
with the radial wave functions $  f_{q \pi n \kappa \Omega}(r)  $
and vector-spherical harmonics
\begin{equation} \label{sph}
 {\cal Y}_{\kappa \Omega}(\hat{r}) = \sum_{m \nu}
 ( l \: m \: s \: \nu \mid  j \: \Omega) Y_{lm}(\hat{r})
 \chi_{s\nu}      
\end{equation}
which are obtained by coupling the orbital angular momentum $  l  $
of the spherical harmonics $  Y_{lm}  $ and the spin $  s=\frac{1}{2}  $
of the spinors $  \chi_{s\nu}  $ to the
total angular momentum $  j  $. The index $  \kappa  $
of the vector-spherical harmonic of Eq.\ (\ref{sph}) specifies
$   j = | \kappa | -\frac{1}{2}   $
with $  l = \kappa-1  $ for $   \kappa > 0  $ and
$   l = -\kappa  $ for $  \kappa < 0  $.
The sum in Eq.\ (\ref{phi}) runs over all $  \kappa  $ with
$  \mid \kappa\mid  \geq \mid \Omega\mid  +\frac{1}{2}  $ where
$  \kappa \in \{1,-2,3, -4, \dots \}  $
for positive parity states and $  \kappa \in \{-1,2,-3,4,\dots \}  $
for negative parity states. The Schr\"{o}dinger equation for
$  \phi_{q\pi n \Omega}  $
leads to a set of coupled differential equation for the radial wave 
functions
$  f_{q \pi n \kappa \Omega}(r)  $ for all allowed $  \kappa  $. In the 
actual calculation only contributions with
$ | \Omega |  +\frac{1}{2} \leq | \kappa |  \leq 
| \Omega |  +\frac{25}{2}  $
are considered. The radial wave functions
are discretized on a grid with a step size of $  h=0.2  $~fm inside an
interval $  [0,R]  $ with maximum radius $  R=(1.25A^{1/3}+12)  $~fm 
for a nucleus
with $  A  $ nucleons. Derivatives are represented by five-point 
formulas.

Particle densities $  \varrho_{q}  $, kinetic densities $  \tau_{q}  $
and spin-current densities $  \vec{J}_{q}  $
appearing in the Skyrme-Hartree-Fock potentials and the energy density
are easily calculated from the single particle wave functions
$  \phi_{q\pi n \Omega}  $.
E.g., proton and neutron single particle densities are given by
the multipole expansion
\begin{equation}
 \varrho_{q}(\vec{r}) = \sum_{L} \varrho_{qL}(r)
 Y_{L0}(\hat{r})
\end{equation}
where
\begin{equation}
 \varrho_{qL}(r) = \sum_{\pi n \Omega } \frac{w_{q \pi n \Omega}}{r^{2}}
 \sum_{\kappa \kappa^{\prime}}
 C_{\kappa \kappa^{\prime}}^{L \Omega}
  f_{q \pi n \kappa \Omega}^{\ast} f_{q \pi n \kappa^{\prime} \Omega}
\end{equation}
with coefficients
\begin{equation}
 C_{\kappa \kappa^{\prime}}^{L\Omega}  =
 \int \,d\Omega \: {\cal Y}_{\kappa \Omega}^{\dagger} Y_{L0}
 {\cal Y}_{\kappa^{\prime}\Omega}
\end{equation}
The occupation probabilities $  w_{q\pi n \Omega}  $
in each state are determined by the BCS calculation.
Only even values of $  L  $ appear in the sum of Eq.\ (3) and 
contributions
$  0 \leq L \leq 10  $ are considered in the calculation.
Non-radial contributions of the spin-current density are neglected in 
the calculation.

The binding energy of nucleus with $  A  $ nucleons and $  Z  $ protons
in its ground state is calculated from
\begin{equation}
 BE(A,Z) = -(E_{mf} + E_{pair} - E_{cm} - E_{rot})
\end{equation}
with the mean-field contribution
\begin{equation}
 E_{mf} = \int d^{3}r \: H(\vec{r})
\end{equation}
which is obtained by integrating the Skyrme-Hartree-Fock
energy density $  H(\vec{r})  $ over
the spatial coordinates. The pairing energy in the BCS approach is 
given by
\begin{equation}
 E_{pair} =  - \sum_{q} \frac{G_{q}}{4} \left( \sum_{\pi n \Omega}
 \sqrt{w_{q \pi n \Omega}(1-w_{q \pi n \Omega})} \right)^{2} \: .
\end{equation}
The pairing strength is
$  G_{+1}=1.9/\sqrt{A}  $~MeV for protons and 
$  G_{-1}=1.2/\sqrt{A}  $~MeV
for neutrons, respectively. These values were obtained from a
fit to experimental pairing gaps of $  N=146  $ isotones and
$  Z=92  $ isotopes.
Only bound states are considered in the determination
of the occupation probabilities $  w_{q \pi n \Omega}  $ in the BCS 
calculation. The correction for the center-of-mass motion
\begin{equation}
 E_{cm} = \frac{3}{4} \left( 45 A^{-\frac{1}{3}} - 25 
 A^{-\frac{2}{3}}\right) \:
 \mbox{MeV}
\end{equation}
is the same harmonic oscillator approximation
as for spherical nuclei in the SKX parametrization.
The rotational correction is approximated by
\begin{equation}
 E_{rot} = \frac{\langle \psi | J_{x}^{2} | \psi \rangle}{2 
 {\cal I}_{x}}
\end{equation}
where $  \psi  $ is the many-body wave function of the nucleus in the 
BCS ground state.
The moment of inertia $  {\cal I}_{x}  $ for the rotation around the $  
x  $-axis is calculated in the cranking model.

\section{Results}
\label{sec3}

\begin{figure}[ht]
\centerline{\includegraphics[scale=0.6]{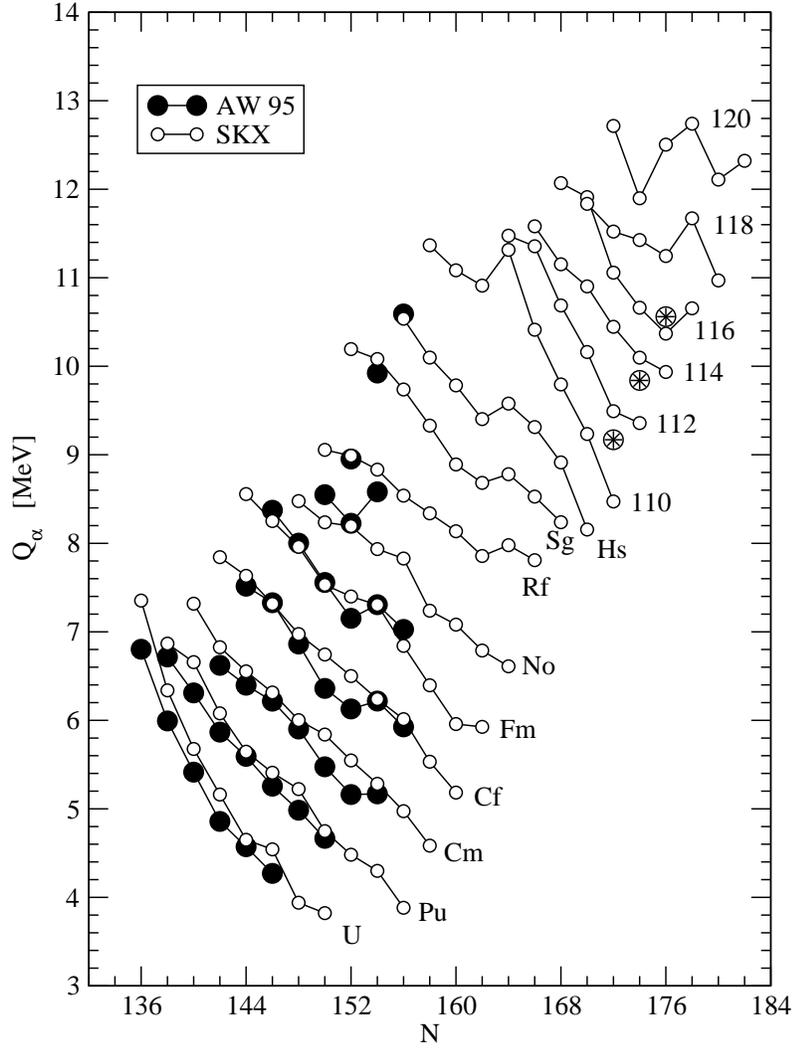}}
\caption{\label{fig2}
Q-value for $  \alpha  $-decay as a function of the
neutron number $  N  $ connected by lines for the given
$  Z  $ values. The experimental values are shown by
the solid circles. The results from the
SKX deformed HF calculations are given by the open circles.}
\end{figure}

The deformed HF calculations give the binding energies for the
nuclei above $^{208}$Pb. From these we calculate the alpha-decay
Q-value, $  Q_{\alpha} = BE(A-4,Z-2)+BE(4,2)-BE(A,Z)  $, where the
experimental value of $  BE(4,2) =  28.30$~MeV is used for the alpha
particle.
The results for alpha-decay Q-values for even-even nuclei are
shown in Fig.\ \ref{fig2}. 
The purpose is to compare with the measured Q-values
as well as to compare with the results based on the SLy4 interaction
shown in Fig.\ 1 of Ref.\ \cite{Cwi99}. Comparison of the two
figures shows that the results for SLy4 and SKX are remarkably
similar even though they are based upon Skyrme parameter sets
which are determined completely independently.
Both show good overall agreement with experimental Q-values \cite{Aud95}
to within a rms deviation of a few hundred keV, with
the exception of a dip in the theoretical Q-values around $  N=152  $
which is not present in experiment. The largest deviation for SKX is
for the Q-value for $^{256}$No at $  N=154  $. 
We also show in Fig.\ \ref{fig2}
the comparison with experimental $  Q_{\alpha}   $ values from
the suggested placement of the $  Z=116-114-112  $ decay chain 
\cite{z116}.
These also agree well with theory.

\begin{figure}[ht]
\centerline{\includegraphics[scale=0.6]{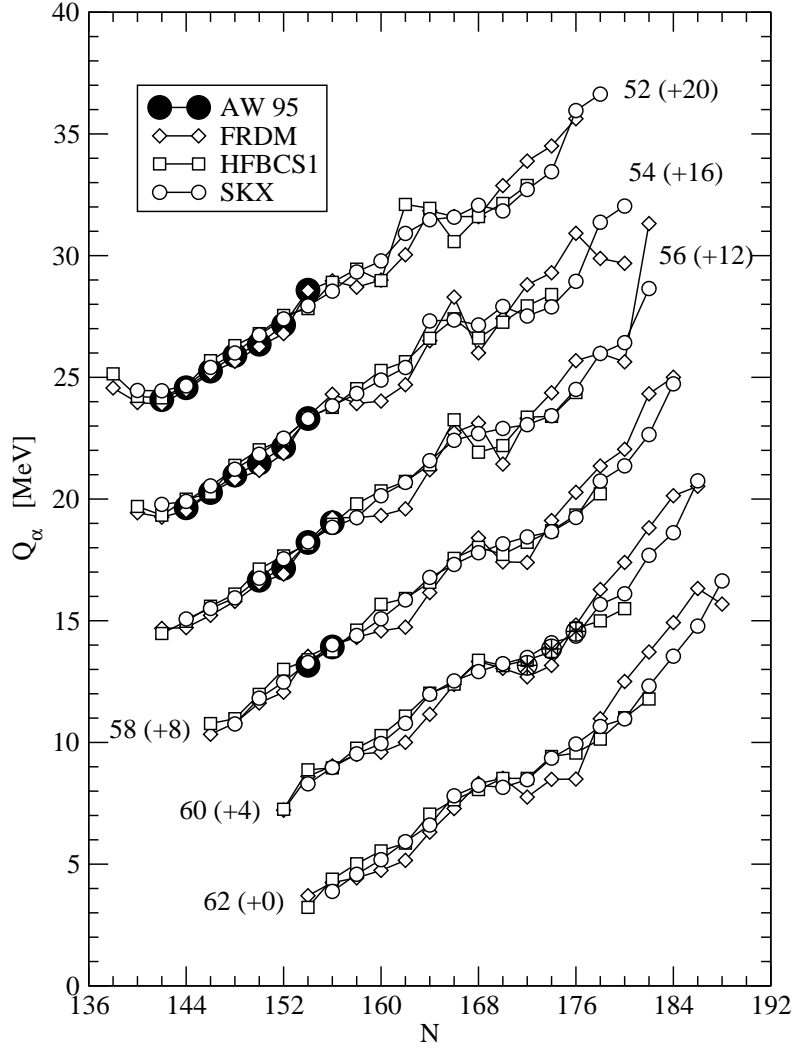}}
\caption{\label{fig3}
Q-value for $  \alpha  $-decay
as a function of the neutron number $  N  $ for even-even nuclei
located on the solid lines of Fig.\ \ref{fig1}.
Predictions from the finite range droplet model (open diamonds: FRDM),
and two Skyrme Hartree-Fock parametrizations
(open squares: MSk7  and  open circles: SKX)
are compared with the experimental data (solid circles).
Different decay chains with
the same value of $  N-Z  $ (indicated by the numbers) are connected
by solid lines and shifted vertically by the amount (in MeV)
shown in parentheses.}
\end{figure}

\begin{figure}[ht]
\centerline{\includegraphics[scale=0.6]{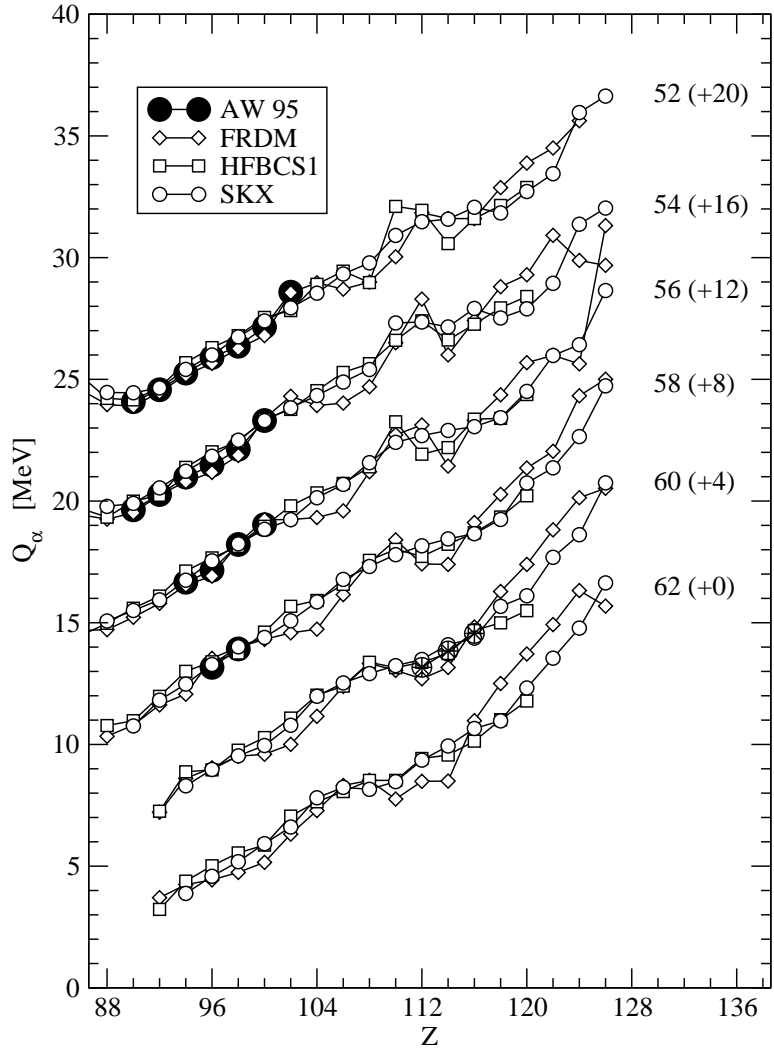}}
\caption{\label{fig4}
Same as Figure \ref{fig3} as a function of $  Z  $.}
\end{figure}

Further results are shown in Fig.\ \ref{fig3} 
as a function of neutron number and in Fig.\ \ref{fig4} 
as a function of proton number. The points are connected
in these figures for a given $  N-Z  $ value in order to emphasize how 
the
Q-values changes in a given decay chain. Comparisons are made to the
finite-range droplet model (FRDM) \cite{Mol95} and to the 
deformed HF-BCS
calculations based on the MSk7 Skyrme interaction \cite{Gor01}.
The results for SKX and MSk7 are very similar
even for the extrapolation to large $  N  $ and $  Z  $. 
The FRDM results
are similar to the HF in the region where data are available but become
more different for the extrapolation to heavier nuclei.

\begin{figure}[ht]
\centerline{\includegraphics[scale=0.6]{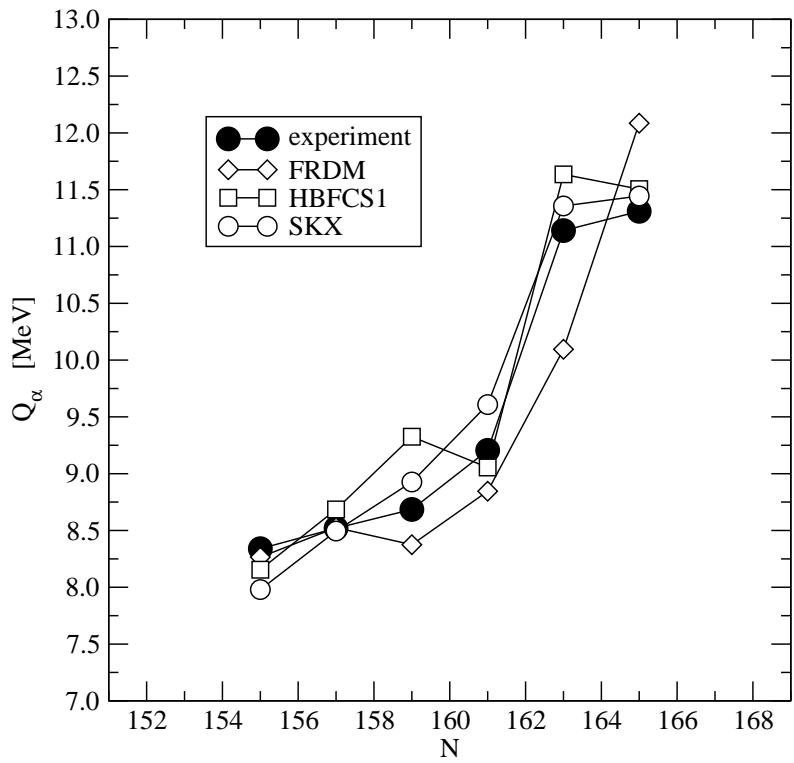}}
\caption{\label{fig5}
Same as Figure \ref{fig3} but for the nuclei with $  N-Z=53  $.}
\end{figure}

Much of the data for the super-heavy nuclei are for 
odd-even decay chains.
These are more difficult to calculate and compare with experiment since
the deformed level density is high and the observed nuclei
may be in isomeric states. These must be considered carefully. For this
paper we compare with the Q-values observed for the recently confirmed
decay chain for $  N-Z=53  $ starting at $^{277}$112 in Fig.\ \ref{fig5}. 
In the calculation
we assume that the nucleus is in its lowest energy deformed
single-particle state. The results are also compared to the FRDM and
MSk7 models. As in Figs.\ \ref{fig3} and \ref{fig4}, 
the SKX and MSk7 results are close
to each other and both are close to experiment, with perhaps SKX being
in best agreement with experiment. The FRDM results do not agree as well
in detail with experiment. In the deformed HF the jump in Q-value 
between $  N=161  $ and $  N=163  $ observed in Fig.\ \ref{fig5} 
comes from a deformed shell
gap at $  N=162  $ and $  Z=108  $. 
These deformed gaps are also found with
the SLy4 interaction \cite{Cwi99}. We note the
semi-empirical shell-model mass approach \cite{Lir02} cannot and does
not take into account these deformed shell gaps and cannot reproduce
any of the fine structure in the Q-value systematics.

\begin{figure}[ht]
\centerline{\includegraphics[scale=0.6]{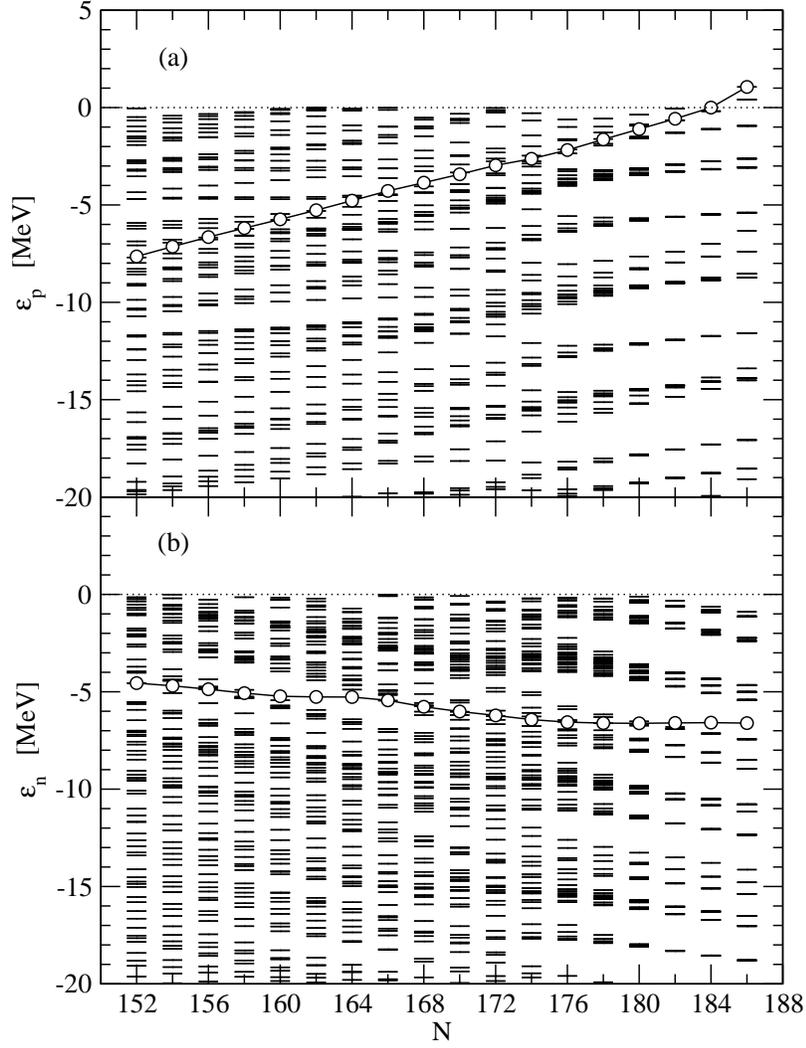}}
\caption{\label{fig6}
Single particle energies of protons (a) and
neutrons (b) above $-$20~MeV in even-even nuclei with $  
N-Z=60  $ in the
Skyrme Hartree-Fock calculation with the SKX parametrization.
The Fermi energies of protons and neutrons
are denoted by open circles.}
\end{figure}

The distribution of the single-particle energies for protons
and neutrons is shown in Fig.\ \ref{fig6} as a function of the neutron
number $  N  $ in even-even nuclei for the $   N-Z=60   $ decay chain.
Nuclei with small $   N   $ are well-deformed and become more and more
spherical with increasing $   N   $. Proton and neutron shell gaps
are readily seen in both spherical and deformed nuclei.
The proton Fermi energy increases smoothly with increasing $   N   $
and becomes positive for the nucleus with $   N = 188   $
which is predicted to be proton-unbound in the
SKX parametrization.
The neutron Fermi energy decreases only slightly
with increasing $   N   $. It crosses the well defined  shell gap
at $   N = 162   $ which was mentioned above.

\begin{figure}[ht]
\centerline{\includegraphics[scale=0.6]{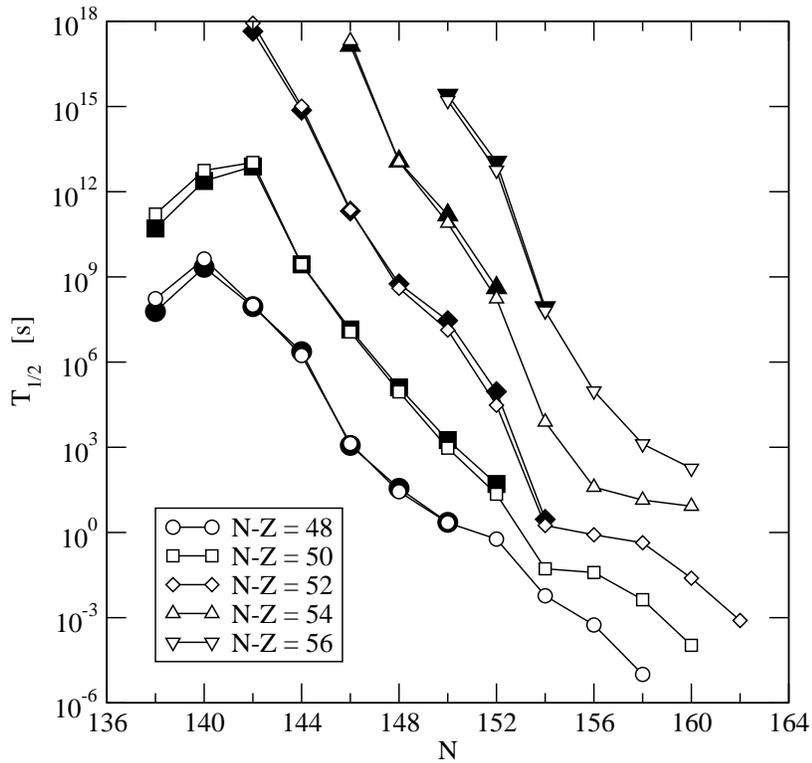}}
\caption{\label{fig7}
Half-lives of even-even nuclei as a function of
the neutron number $  N  $. Open symbols indicate half-lives calculated
with Eq.\ (4) with the experimentally measured
$  Q_{\alpha }  $ values or with $  Q_{\alpha }  $ values from the
Audi-Wapstra mass extrapolation.
Solid symbols denote experimental
half-lives. Decay chains with constant
$  N-Z  $ value are connected by solid lines.}
\end{figure}

The alpha-decay half-life is important for determining how the alpha
decay of super-heavy nuclei competes with fission. The
extrapolated half-lives are also important for choosing
the type of experimental
techniques used for their identification. The main theoretical
uncertainty for the calculation of the assumed $  L=0  $ decays of
even-even nuclei is in the alpha-decay Q-value. To calculate the
half-lives we use the empirical result obtained in \cite{Bro92}
\begin{equation} \label{t12}
 \log_{10}\left[T_{1/2}/\mbox{s}\right] =
 9.54 (Z-2)^{0.6}/\sqrt{Q_{\alpha}/\mbox{MeV}}-51.37  
\end{equation}
In Fig.\ \ref{fig7} we show the half-life calculated from 
Eq.\ (\ref{t12}) and from
the experimental Q-values \cite{Aud95}. The results are compared to the 
experimental half-lives. 
The excellent agreement between experiment and theory
shows that preformation and decay systematics implied by Eq.\ (\ref{t12})
are adequate for a determination of the alpha-decay half-life to within
about a factor of three.

\begin{figure}[ht]
\centerline{\includegraphics[scale=0.6]{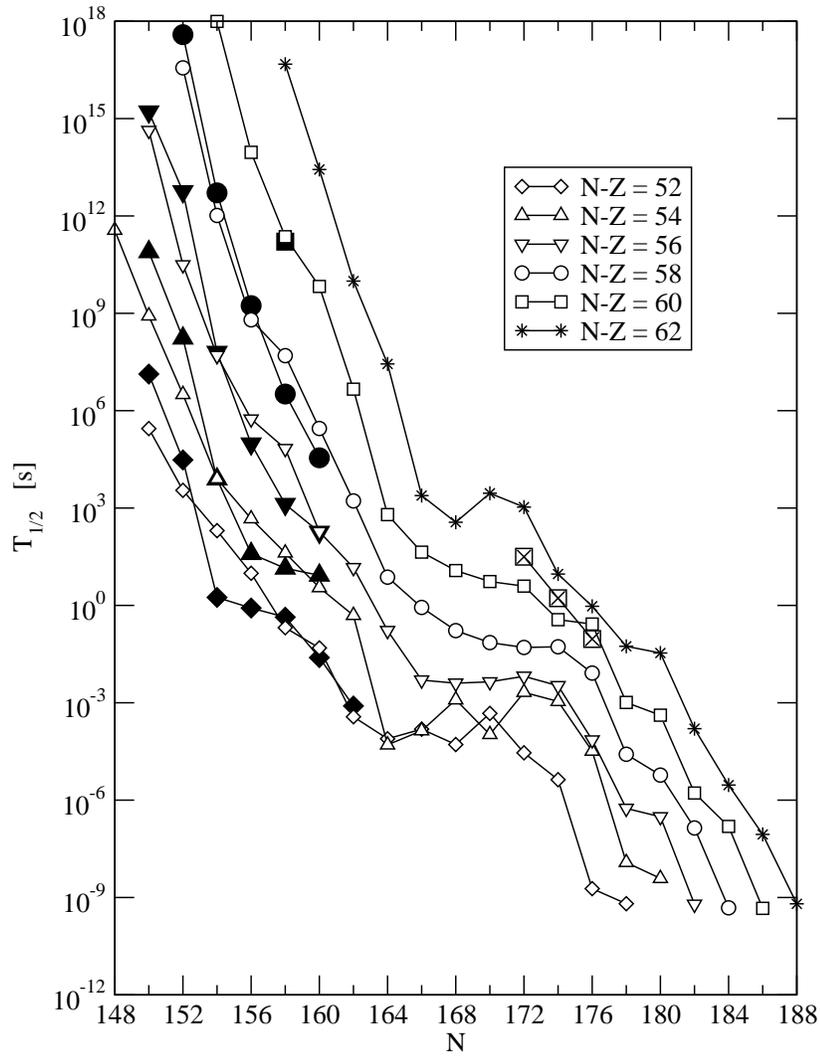}}
\caption{\label{fig8}
Half-lives of even-even nuclei as a function of
the neutron number $ N  $. Open (solid) symbols 
indicate half-lives calculated
with $  Q_{\alpha}  $ values from the SKX parametrization
(experiment). Decay chains with constant
$  N-Z  $ value are connected by solid lines. The experimental
half-lives from the suggested placement of the $  {}^{292}116  $ 
decay chain are shown by the cross-filled boxes.}
\end{figure}

The predictions for the half-lives of heavier nuclei
based upon the theoretical $  Q_{\alpha }  $ values from our
SKX calculations
results are shown in Fig.\ \ref{fig8}. The agreement with experiment is
satisfactory except with near $  N=152  $ where the kink in the
experimental half-lives is not reproduced by the theory.
The experimental half-lives for the
suggested placement of the $  Z=116  $ decay chain \cite{z116}
are also in reasonable agreement with theory.
One observes an island of relative
stability starting at $  N=164  $ where the half-lives for
$  Z\approx 53  $ remain at the msec level or longer
until $  N\approx 174  $ where
they start to become shorter.

\section{Summary}

We have presented a new calculation for the alpha decay
$  Q  $ values for super-heavy nuclei based upon deformed Hartree-Fock
calculations with the SKX Skyrme interaction. A new computational
method is used to carry out axially-symmetric deformed calculations.
Agreement with experimental data 
including the recently observed $  Z=112  $,
$  Z=114  $ and $  Z=116  $ decays is obtained at the rms level of
a few hundred keV. Deformed shell gaps at  $  N=162  $
and $  Z=108  $ lead to jumps in the $  Q  $ values which are
consistent with experiment. The $  Q  $ values have been used to
calculate alpha-decay half-lives which are in reasonable agreement with
theory. Predictions for the $  Q  $ values and half-lives up to the
proton drip line at $  N=184  $ and $  Z=126  $ are made.

\acknowledgments

This work was supported by U.S. National Science Foundation
grant PHY-0070911.

\clearpage

\begin{table}
\caption{\label{tab1}
Single-particle energies for states in $^{208}$Pb and the
rms difference between experiment and theory.}
\begin{tabular}{crrrr}
\hline \hline
      & \multicolumn{4}{c}{single-particle energy (MeV)} \\
     orbit           &   exp    &    SKX   &   SLy4    &  MSk7  \\
\hline
$  \pi 1h_{9/2}  $   &  $-$3.80 &  $-$4.26 &  $-$3.82  & $-$3.47 \\
$  \pi 2._{7/2}  $   &  $-$2.90 &  $-$2.83 &  $-$2.93  & $-$3.19 \\
$  \pi 1i_{13/2} $   &  $-$2.29 &  $-$2.18 &  $-$1.46  & $-$2.62 \\
$  \pi 2f_{5/2}  $   &  $-$0.98 &  $-$0.70 &  $-$0.37  & $-$0.93 \\
\hline
$  \nu 2g_{9/2}  $   &  $-$3.94 &  $-$3.46 &  $-$3.14  & $-$4.05 \\
$  \nu 1i_{11/2} $   &  $-$3.16 &  $-$2.81 &  $-$1.57  & $-$2.02 \\
$  \nu 1j_{15/2} $   &  $-$2.16 &  $-$1.97 &  $-$0.64  & $-$2.53 \\
$  \nu 2g_{7/2}  $   &  $-$1.45 &  $-$1.00 &     0.08  & $-$0.97 \\
\hline
rms & & 0.34 & 1.05 & 0.51 \\
\hline \hline
\end{tabular}
\end{table}

\end{document}